\documentclass[aps,pra,floatfix,twocolumn]{revtex4}
\usepackage{graphicx}
\usepackage{amsmath}
\usepackage{amssymb}
\usepackage{amsfonts}
\usepackage{amsthm}
\usepackage{verbatim}
\usepackage{times}
\usepackage{helvet}
\usepackage{courier}
\usepackage{bm}
\usepackage{url}

\begin{document}
\title{Analysis of self-overlap reveals trade-offs in plankton swimming trajectories}

\author{Giuseppe Bianco$^1$, Patrizio Mariani$^2$, Andre W. Visser$^2$, Maria Grazia Mazzocchi$^3$ and Simone Pigolotti$^4$}

\affiliation{
$^1$Department of Biology, Ecology Building, Lund University, SE-223 62 Lund, Sweden \\ 
$^2$Center for Ocean Life, National Institute for Aquatic Resources, Technical University of Denmark, Kavalerg{\aa}rden 6, 2920 Charlottenlund, Denmark \\
$^3$Stazione Zoologica Anton Dohrn, Villa Comunale, 80121 Napoli, Italy \\
$^4$Departament de Fisica i Enginyeria Nuclear, Universitat Politecnica de Catalunya Edifici GAIA, Rambla Sant Nebridi 22, 08222 Terrassa, Barcelona, Spain}


\begin{abstract}
Movement is a fundamental behaviour of organisms that brings about beneficial encounters with resources and mates, but at the same time exposes the organism to dangerous encounters with predators. The movement patterns adopted by organisms should reflect a balance between these contrasting processes. This trade-off can be hypothesized as being evident in the behaviour of plankton, which inhabit a dilute 3D environment with few refuges or orienting landmarks. We present an analysis of the swimming path geometries based on a volumetric Monte Carlo sampling approach, which is particularly adept at revealing such trade-offs by measuring the self-overlap of the trajectories. Application of this method to experimentally measured trajectories reveals that swimming patterns in copepods are shaped to efficiently explore volumes at small scales, while achieving a large overlap at larger scales. Regularities in the observed trajectories make the transition between these two regimes always sharper than in randomized trajectories or as predicted by random walk theory. Thus real trajectories present a stronger separation between exploration for food and exposure to predators. The specific scale and features of this transition depend on species, gender, and local environmental conditions, pointing at adaptation to state and stage dependent evolutionary trade-offs.
\end{abstract}

\keywords{searching strategy, encounter rate, zooplankton, self-overlap}

\maketitle


\section{Introduction}
Animal movement is a fundamental process in ecology that affects interactions between individuals, their mates, prey and predators, and ultimately structures many of the properties of an ecosystem \cite{Nathan2008}. Movement has profound implications on the ecology and evolution of large classes of organisms including microorganisms, plants and animals. A key aspect directly regulated by organisms' motion is the encounter process, the fundamental currency of all trophic and reproductive interactions. Migrations, food or nutrient acquisition, mating, and hazardous meetings with predators, are all examples of dynamics regulated by encounters and hence directly controlled by the motion patterns expressed by organisms both at individual \cite{Kiorboe2008book, Bazazi2011} and collective level \cite{Couzin2005}.

It is relatively well understood how encounter processes depend on the relative velocities between individuals and on the perception distance of the searching animal \cite{Gerritsen1977}. Path geometry too is thought to play an important role, although a characterization of its impact remains elusive.  Examples of theoretical ideas and proposed metrics of path geometry include net-to-gross displacement ratio \cite{Buskey1984}, searching efficiency \cite{Bundy1993, Uttieri2008}, the statistics of run lengths \cite{Viswanathan1996, Bartumeus2003}, the diffusive-ballistic transition \cite{Visser2006}, composite random walks and patchiness \cite{Benhamou2007}, random walks, spirals and the Wiener sausage \cite{Levandowsky1988, Visser2007}, fractal dimension and space filledness \cite{Schmitt2001}. While this list is by no means exhaustive, it illustrates the wide variety of approaches adopted to assess the ecological relevance of movement patterns.

One of the main goals of the theoretical frameworks above has been to determine how a given motion pattern contributes to the search efficiency of an organism. However, while the imperative for motion is primarily related to the search for resources, it is also fraught with risk. As well as incurring an energy cost, a moving animal becomes more conspicuous and runs the risk of blundering into a predator. One should expect that movement behaviour would balance the aforementioned risks against the necessity of searching for food \cite{Visser2006, Rosier2011}. It is our contention that ecologically meaningful analyses of movement should try to infer features of the underlying trade-off from path geometry data.

While path geometry is the outward projection of swimming behaviour, its relevance for the organism is the volume the path sweeps out in three-dimensional space and the number of prey and predators it encounters. This volume is inherently scale dependent, depending on the distance from the path within which detection can take place. The detection distance can vary for different classes of encounter; it may be larger for large prey compared to small prey, and even larger for detection by a predator.  With respect to the fitness trade-off, two relevant detection distances can be identified: the one at which the organism can detect its prey in the case of search, and that at which it can be detected by a predator.  

Although the problem of motility, path geometry and trade-offs is common to many life forms, we discuss this here in the context of plankton. Plankton inhabit a dilute three-dimensional environment, where there are few orienting cues, perception is limited, and the distances between individuals (both predators and prey) are relatively large. In such environment, movement assumes an important role, and while their fluid environment imposes movement in its own via currents and turbulence, many planktonic organisms, even algae, have locomotion ability. We thus focus on plankton motion in a quiescent fluid; a condition easily reproducible in the laboratory, and occurring in many natural situations.

Planktonic organisms have been reported to swim with very tortuous paths. For high search efficiency this path should have very low overlap and correspond to a large swept volume for radius equal to the species detection radius. At the same time, a large overlap at a radius corresponding to the typical range at which predators can detect its prey would lead to a low exposure to mortality risk. In plankton, the detection distance by a predator generally exceeds the detection distance for prey by about an order of magnitude \cite{Hansen1997} leading at small overlap (i.e. high encounter) with prey and large overlap (low encounter) with predators.

In this paper, we propose a fast and accurate way of directly estimating self-overlap of zooplankton trajectories by means of a Monte Carlo scheme. We apply the method to a set of three-dimensional copepod trajectories and show that this measure highlights the change in the trade-offs for different species and gender, different swimming activities, and at varying food conditions.

\section{Results}

\subsection{Properties of self-overlap}

\begin{figure}
\includegraphics[width=84mm]{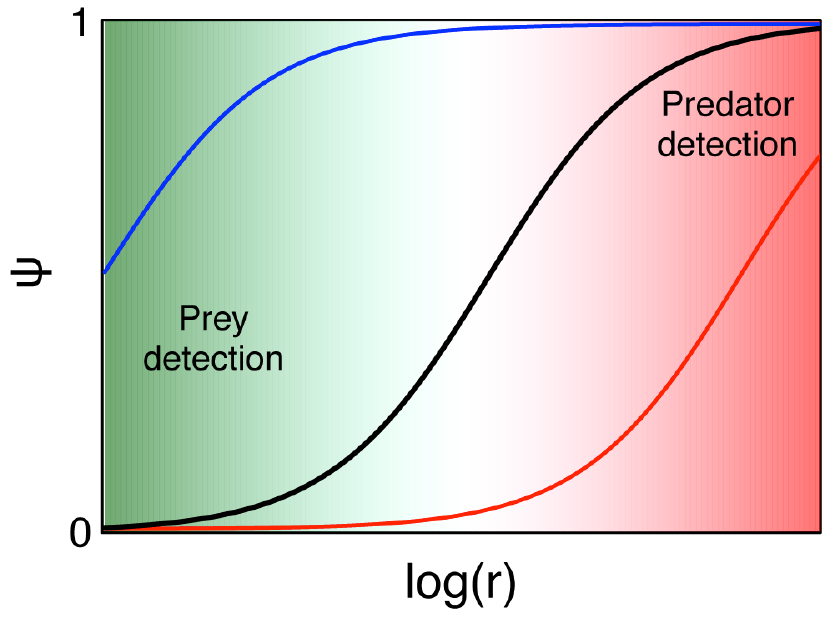}
\caption{Examples of the self-overlap $\psi$ as function of detection radius $r$. At small $r$ the probability of self-overlap is low  while for larger $r$ the self-overlap is higher. Green and red areas are prey and predator detection ranges, respectively. The transition of $\psi$ from low to high values depends on path geometry. In the case of a random walk (black curve, see Eq. \ref{RW} for the analytical solution) the transition is smooth. However, real organisms' trajectories can be shaped to be more exposed to predation or shaped to reduce predation risk as hypothesised with the red and blue lines, respectively.}
\label{fig0}
\end{figure}

We want to calculate the volume $V(r)$ swept by a trajectory characterized by a detection distance $r$.
We call $L$ the total length of the path and assume for simplicity that the perception distance is isotropic in all directions. Then, taking into account the start and end points of the trajectory, the maximum volume swept out corresponds to the one of a straight line:
\begin{equation}\label{V_max}
V_{max}(r)=\frac{4}{3}\pi r^3 +\pi r^2L
\end{equation}
Depending on the tortuosity of the path compared to $r$, a varying portion of the swept volume will be self-overlapping so that in general $V(r) \le V_{max}(r)$. It is then natural to define the {\em scale-dependent self-overlap}
\begin{equation}\label{def_selfover}
\psi(r)=[1-V(r)/V_{max}(r)]
\end{equation}
By definition, the admissible values of $\psi(r)$ are between 0 and 1, with $\psi(r)=0$ when the trajectory never intersects itself, as for example in the case of a rectilinear motion, and $\psi(r)=1$ when the trajectory is always visiting portions of space that have been previously scanned. Moreover, $\psi(r)$ is a strictly increasing function of $r$ and the probability of self-overlapping is very small at small scales, i.e., $r \rightarrow 0$, $\psi(r) \rightarrow 0$, while it is higher at larger scales, i.e., $r \rightarrow \infty$, $\psi(r) \rightarrow 1$ as sketched in Figure~\ref{fig0}.

\begin{figure}
\includegraphics[width=84mm]{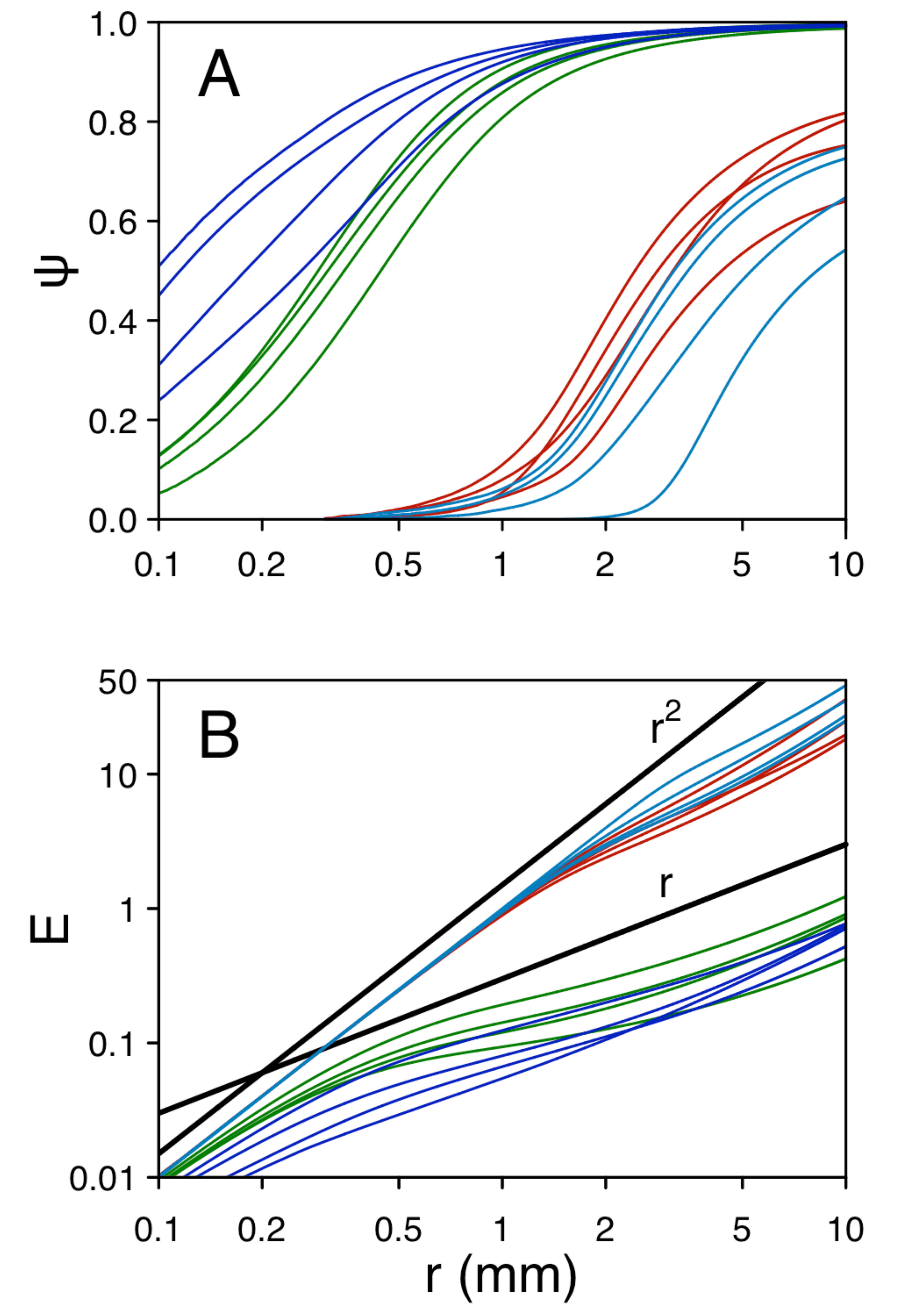}
\caption{(A) Self-overlap $\psi$ of trajectories shown in Fig.~\ref{fig1} and corresponding (B) encounter rates $E$ estimated from Eq.~\ref{enc_formula} with $v=\rho=1$. Also reported $r^2$ and $r$ for comparison with ballistic and random trajectories, respectively.}
\label{fig2}
\end{figure}

Measures of $V(r)$ for real trajectories cannot be obtained with simple analytical procedures. For this reason we employed a Monte Carlo algorithm, which is a computationally efficient tool, to estimate the volumes in Eq. \ref{def_selfover} (see Method section). The self-overlap measured for a subset of zooplankton trajectories in our database shows clearly the general patterns described above (Fig.~\ref{fig2}A). Nevertheless $\psi(r)$ appears to be highly variable especially in the transition from low to high values (Fig.~\ref{fig2}A).

The variability in self-overlap is ecologically significant as it contributes to determining the encounter rates with prey and predators. The average encounter rate of an individual, taking into account the self-overlap, can be estimated as:
\begin{equation}\label{enc_formula}
E(r)=\pi r^2 v \rho (1-\psi(r))
\end{equation}
where $v$ is the swimming speed, $\rho$ the number density of encounter items (prey, mates or predators), and $r$ the considered detection distance. For simplicity Eq.~\ref{enc_formula} neglects any movement of encounter partners and the possibility of renewal of resources and assumes $100\%$ clearance efficiency (i.e., all items in the path are captured). Further, $E(r)$ must be considered as the long-term encounter rate, i.e. after a sufficient portion of space has been explored. Under these assumptions, and since large self-overlap implies that on average the organism is visiting regions of space that have been already explored, we conclude that at large $\psi$ a consequent reduction of the encounter probability is obtained. 

The shape of the curve $E(r)$ contains important information about the range of possible encounters of an individual and potentially about its fitness, which is enhanced by a high encounter rate with prey and a low encounter rate with predators \cite{Visser2007}. Even taking into account self-overlap, $E(r)$ is always an increasing function of $r$, as larger detection radii always lead to larger encounter rates.

In the limiting case of a ballistic trajectory, $\psi(r)=0$, $E(r)$ is simply proportional to $r^2$, while $E(r)\propto r$ for a three-dimensional random walk \cite{Rednerbook}. A correlated random walk presents both regimes: $E(r)$ increases as $r^2$ at small scales and as $r$ at larger scales, with a crossover length set by the correlation length scale \cite{Visser2006}. The direct estimate of the encounter rate (Eq.~\ref{enc_formula}) for the self-overlap presented above (Fig.~\ref{fig2}A) reveals that, depending on the specific trajectory, the function $E(r)$ may present a variety of slopes and regime changes (Fig.~\ref{fig2}B). This is also a consequence of the complex geometry of the trajectories, which is not necessarily captured by a correlated random walk. Although the Eq. \ref{enc_formula} require extra assumptions (see above), the non-trivial regime changes of $E(r)$ is determined only by the function $\psi(r)$, which is an assumption-free measure. For this reason we will concentrate in the following on the self-overlap function only.


\subsection{Analysis of zooplankton trajectories}

\begin{figure}
\includegraphics[width=84mm]{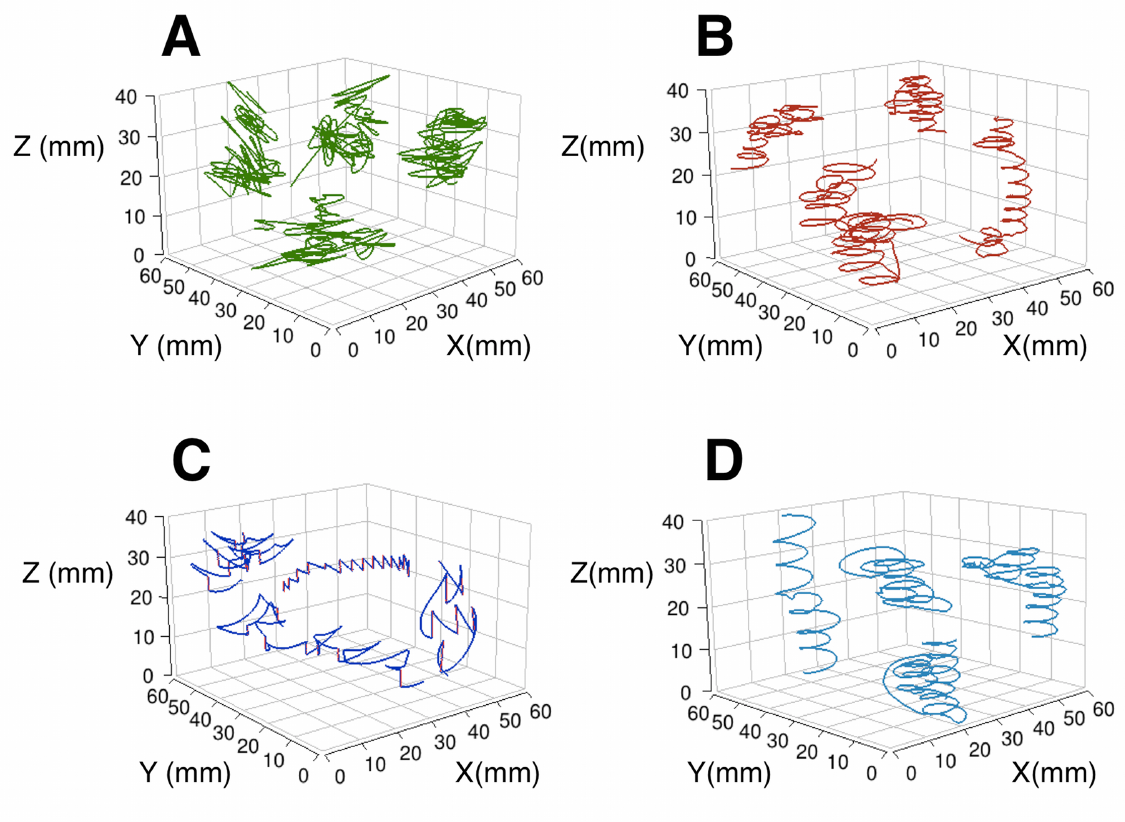}
\caption{Example of motion patterns analysed in the work. Four 1-minute trajectories in each panel: (A) \textit{Clausocalanus furcatus} fast cruising pattern, (B) \textit{Temora stylifera} female, (C) \textit{C. furcatus} swim-and-sink pattern (sink in red), (D) \textit{T. stylifera} male.}
\label{fig1}
\end{figure}

The motion behaviour of the marine calanoid copepods \textit{Clausocalanus furcatus} and \textit{Temora stylifera} show a variety of swimming patterns (Fig.~\ref{fig1}). \textit{C. furcatus} females employ different swimming behaviours depending on prey availability, showing continuous, highly convoluted and fast swimming when food is present (Fig.~\ref{fig1}A), and a set of regular geometric patterns with alternating swimming and sinking phases under food depleted conditions (Fig.~\ref{fig1}C) \cite{Bianco2013}.  In contrast, \textit{T. stylifera}, both females and males, perform only a cruising behaviour with regular loops in the horizontal plane (Fig.~\ref{fig1}B and D).

\begin{figure}
\includegraphics[width=84mm]{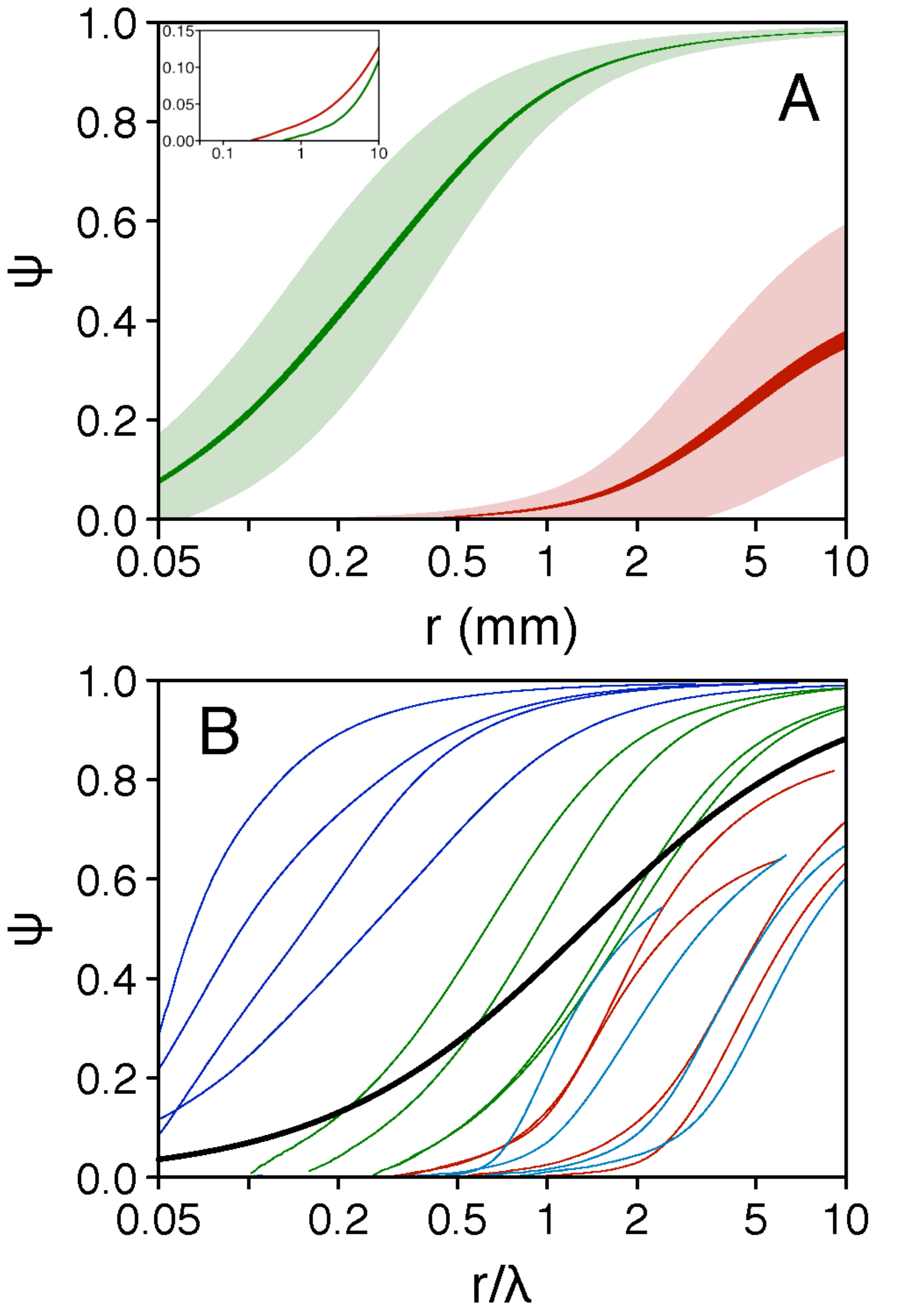}
\caption{Self-overlap of different motion patterns: (A) \textit{Clausocalanus furcatus} (green, N=320) and \textit{Temora stylifera} (red, N=192). Solid area: mean $\pm$ SEM; shaded area: SD. Also reported in the inlet the mean $\psi$ for randomized trajectories for both species. (B) \textit{C. furcatus} (green), \textit{T. stylifera} female (red) \textit{C. furcatus} with sink (dark blue), \textit{T. stylifera} male (light blue). We use the scaling $r / \lambda$ where $\lambda$ is the estimated correlation length scale for each trajectory. Also shown the expected overlap $\psi=3r/(3r+4\lambda)$ for Brownian motion (black line).}
\label{fig3}
\end{figure}

The two species have similar size but quite distinct detection radii ($r \approx 0.2$ $mm$ for \textit{C. furcatus } and $r \approx 1.4$ $mm$ for \textit{T. stylifera}, see Methods section) and have different swimming patterns that can also affect encounter rates with both predators and preys. The analysis of all the trajectories in the database shows a consistently larger self-overlap for \textit{C. furcatus} than for \textit{T. stylifera} (Fig.~\ref{fig3}A). In particular, at a distance of $r = 1$ $mm$, \textit{C. furcatus} has a self-overlap approaching $100\%$ while \textit{T. stylifera} only $5\%$. For both species the standard deviation of individual tracks accounts for $<29.4\%$ on the measured $\psi$ at $r$ typical of the encounter with the prey. It is worth noting that the sharp transition in self-overlap of \textit{C. furcatus} occurs at length scales typical of the transition between the encounters with prey and the encounters with predators (Fig.~\ref{fig3}A). Indeed this species is efficiently scanning the water (low self-overlap) at a distance up to $r=0.2$ $mm$ while it becomes quickly less efficient at larger scales. At $r=1.0$ $mm$, a self-overlap of $\sim100\%$ corresponds to an efficient sheltering and a significant reduction of the encounter probabilities with potential predators.

\subsection{Experimental trajectories and deviation from random walk}

To assess whether the self-overlap curves reflect the presence of regular patterns in the observed trajectories, we compared the experimental dataset with a ''randomized'' counterpart. From each trajectory, a randomized version was generated by measuring all the segments between consecutive sample points, and all angles between consecutive segments. A new trajectory was then generated by randomly reassembling the segments and the angles. Clearly, the randomized trajectory have similar statistical features compared to the original one (including statistics of run lengths, correlation length and statistics of turning angles), but all regularities and large scale structures are lost. The comparison between real and randomized trajectories (Fig.~\ref{fig3}A) shows that (i) randomized trajectories tend to have a much smoother transition between the regime in which the self-overlap is small, and (ii) the self-overlap of randomized trajectories is much lower than the one of real trajectories at the maximum radius considered. 

It is also interesting to compare the experimental curves with that of a Brownian motion, which is often used as a model for animal motion. Self-overlap of a Brownian motion can be phenomenologically estimated as
\begin{equation}\label{RW}
\psi \sim \frac{3 \tilde{r}}{3 \tilde{r} + 4 }
\end{equation}
where we introduced the rescaled radius $\tilde{r}=r/\lambda$, being $\lambda$ the correlation length of the Brownian motion \cite{Visser2007} (see Harris \cite{Harris1982} for a more extensive treatment). Such a comparison is presented in (Fig.~\ref{fig3}B), where the experimental self-overlap curves in Fig.~\ref{fig2}A are rescaled with their correlation length, and the black line is the theoretical prediction. As a consequence of the regular patterns in real trajectories, the transition from low to high self-overlap is not always captured by the correlation length scale, but varies about one order of magnitude greater or smaller than $\lambda$. Moreover a key feature of all curves is that the transition between the two regimes is much sharper than for a random walk. 

\subsection{Sensitivity to environmental conditions}

\begin{figure}
\includegraphics[width=84mm]{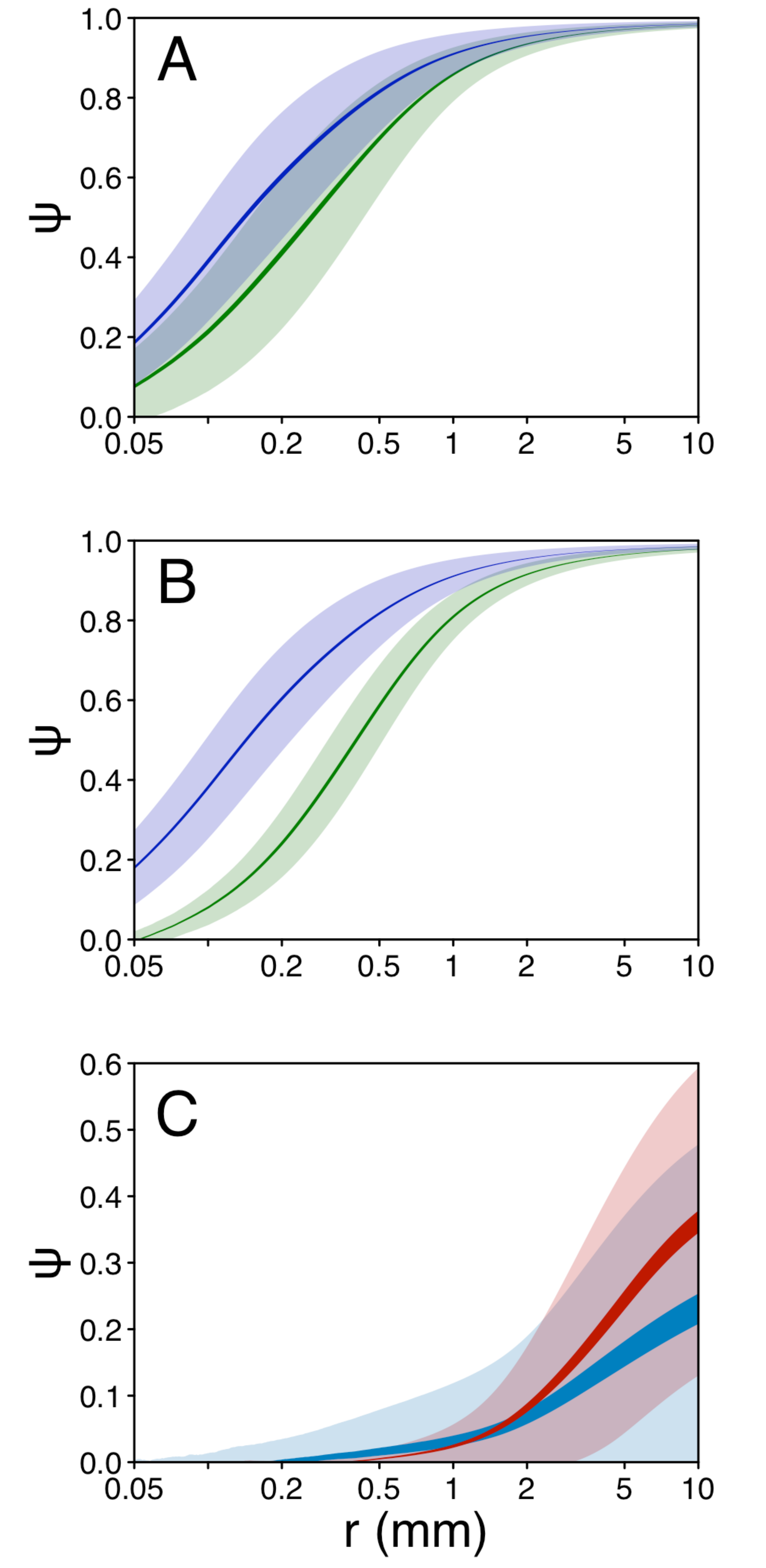}
\caption{Self-overlap for different food, swimming mode, and gender. \textit{Clausocalanus furcatus} in (A) different trophic conditions: green, presence of food (N=320); purple, no food (N=430), and (B) different swimming patterns: green, fast cruising (N=222); purple, swim-and-sink (N=528). \textit{Temora stylifera} (C) different gender: female (red, N=192) and male (blue, N=117). Solid area: mean $\pm$ SEM; shaded area: SD.}
\label{fig4}
\end{figure}

Local environmental conditions, such as food quantity and quality or population structure, are likely to affect swimming behaviour. When \textit{C. furcatus} is exposed to depleted food conditions for a few hours, it drastically reduces its activity (Fig.~\ref{fig1}C), increasing the self-overlap (Fig.~\ref{fig4}A) and reducing the average swimming speed \cite{Bianco2013}. Interestingly, this change in the kinematics is often correlated to a change in the swimming patterns, going from a fast cruising to a swim-and-sink behaviour, hence consistently increasing the amount of volume overlap for all the trajectories (Fig.~\ref{fig4}B). The reduction of motion during unfavourable food conditions is consistent with both observations on other copepod species \cite{Paffenhofer2007} and theoretical considerations about zooplankton fitness optimization \cite{Visser2006}. Feeding under risk of predation is dangerous and when food is scarce the pay-off for a low motility (low risk) strategy is higher than a bolder (less safe) one. In particular at $r=1$ $mm$ a behavioural switch from cruising to a swim-and-sink patterns allows a significant increase in $\psi$ ($\sim15\%$), which act in reducing the encounter probabilities with predators predicted by Eq.~\ref{enc_formula}.

It has been observed that gender can affect swimming behaviour in copepod species \cite{doall1998, dur2010}, with males constrained by the need of finding a mate and showing a bolder motility \cite{Kiorboe2008}. On the other hand, females might show a more risk adverse behavior, likely in the attempt to secure their reproductive output. While there is no significant difference in the shape of trajectories displayed by females and males of \textit{T. stylifera} (Fig.~\ref{fig1}B and D, respectively), their respective self-overlap analyses do show distinct patterns. Indeed males show a consistently lower self-overlap than females (Fig.~\ref{fig4}C). Moreover the rate of change of the self-overlap at large radius ($r > 1$ $mm$) is always positive but significantly higher for females than for males (59\% in average on the 1--10 $mm$ interval).

\section{Discussion}

\subsection{Biological relevance of self-overlap}

The motion of animals often display intriguing non-random features. It is not straightforward assessing how these features contribute to the fitness of the animal, and how they may change in response to changing environmental conditions. In this study we describe a method to quantify the effective volume explored by a moving animal by measuring the self-overlap associated with individual trajectories. This measure is a characteristic of the geometry of the trajectory itself and is provided as a function of a relevant length scale, which can be related to either the animal perception range or to that of its predators. Moreover, it can be used in the calculation of encounter rate when specific speeds and detection distances are assumed (Eq. \ref{enc_formula}).

Applications of the self-overlap to zooplankton trajectories show that the features of path geometry can be related to biologically relevant properties, such as gender, or to environmental conditions, such as different food concentrations. Interestingly, the method has shown to not be influenced by the size of the organisms under study. The $20\%$ of self-overlap occurred at $r=0.2$~$mm$ for \textit{C. furcatus}, at $r=4.0$~$mm$ for \textit{T. stylifera} female, and at $r=7.2$ $mm$ for \textit{T. stylifera} male.  Such differences in self-overlap could not be explained by differences in body size, with species length difference $<50\%$ and no length difference in \textit{T. stylifera} genders (see Methods).

Unlike other metrics that have been previously used to characterize swimming tracks, the self-overlap is an assumptions free measure of the effective volume explored by moving animals. When compared to estimates of the correlation length scale $\lambda$ (Table~\ref{tab2}), the self-overlap proves to be more sensitive to changes in environmental conditions. Indeed, while both methods lead to similar conclusions regarding swimming behaviour in contrasting food environments or between genders, the self-overlap values always display lower standard deviation for the cases we analysed. Moreover, when animals' swimming behaviour includes a sinking phase, the corresponding correlation length scale is larger. This might erroneously suggest that sinking is a more efficient behaviour than cruising for encountering food and predators. The analyses of self-overlap clearly show that this is not the case. In fact, although the swimming is correlated on larger scales, swim-and-sink trajectories reveal that organisms were repeatedly exploring the same volume of water and the motion is then characterized by larger self-overlap and less efficient encounter rates. This result combined with the low speed observed during the swim-and-sink pattern \cite{Bianco2013} would drastically reduce the encounter rate predicted by Eq.~\ref{enc_formula}.

\begin{widetext}
\begin{table}
\caption{Statistics of correlation length scale ($\lambda$) of both species obtained with the fitting of equation 7 in Visser \& Ki{\o}rboe \cite{Visser2006}. Values in parenthesis are obtained removing trajectories with $\lambda$ higher than the gross distance travelled.}
\begin{tabular}{rcccccc}
\hline
\it{species} & \it{gender} & \it{food} & \it{sink} & \it{median (mm)} & \it{mean (mm)} & \it{sd (mm)} \\
\hline
\it{C. furcatus} 	& Female & Yes & -   & 0.6 (0.6) &  1.4  (1.3) &   2.4  (1.8) \\
 						 	&        & No  & -   & 1.5 (1.5) &  2.7  (2.6) &   4.1  (3.8) \\
 						 	&        & -   & No  & 0.6 (0.6) &  1.4  (1.4) &   2.9  (2.9) \\
 						 	&        & -   & Yes & 1.3 (1.3) &  2.2  (2.1) &   3.6  (3.1) \\
\it{T. stylifera} 		& Female & Yes & -   & 2.6 (2.5) & 10.6  (6.5) &  27.4 (12.9) \\
 						 	& Male   & Yes  & -   & 5.7 (4.8) & 37.9 (11.1) & 177.4 (14.9) \\
\hline
\end{tabular}
\label{tab2}
\end{table}
\end{widetext}

We applied our method to zooplankton primarily because it can be considered as an ideal case to test our method. Indeed, planktonic organisms live in a relatively homogeneous and diluted environment and presumably possess limited information about the spatial distribution of food and predators. While the swimming pattern may change with the presence of predators, the organism can not really know if predators are present or not, thus any searching strategy in plankton should involve some degree of sheltering. 

Although copepod tracking can be performed in the laboratory under simulated turbulence \cite{Yen2008}, we addressed the simple case of a quiescent fluid that is not an uncommon situation in pelagic ecosystems. In the study site, during the season of peak abundance of the two copepod species under study (i.e., late summer-early autumn), the water column is stabilised by thermal stratification \cite{Ribera2004} which implies a relatively low turbulence levels.

Animals with more complex perception mechanisms could use additional information to adjust their movement strategy, by changing features of their trajectories. However, we expect that the basic features, i.e. efficient exploration at small scales and avoiding exposure at large scales, should be ubiquitous among many animal species \cite{Visser2006}. The measure of self-overlap could then be applied to characterize trajectories of species across scales and environments to address the trade-offs of searching strategies in aquatic organisms (e.g., fishes, turtles, mammals), in aerial animals (e.g., insects and birds) as well as in the terrestrial habitats.

\subsection{Ecological and evolutionary implications}

Changes in motion patterns usually result in different scale-dependent self-overlap. This has relevant effects on instantaneous encounter probabilities and thus fitness of zooplankton on the long term. Low self-overlap increases the foraging efficiency of the animals at short length scales while it can yield enhanced predation risk and therefore lower zooplankton fitness on larger length scales. We suggest that the self-overlap analysis is an effective way to assess the evolutionary significance of the path geometry.

Numerical modelling have shown that uncorrelated motions are more exposed to predation risk  \cite{uttieri2010} suggesting high cost for zooplankton individual's fitness. Our comparison of real trajectories with their randomised counterparts supports the view that evolution selects movement patterns characterized by an abrupt transition in the self-overlap: from a very small value at small scale, hence efficient volume exploration for prey and mates, to large overlap at larger scales, hence reduced exposure to predators (Fig.~\ref{fig3}A). Such a conclusion is supported by the comparison with the expected self-overlap of a correlated Brownian motion, presented in Fig.~\ref{fig3}B. As observed with the randomized trajectories, the transition from "efficient search" to "reduced risk" for Brownian motion is less abrupt than for real trajectories, indicating some higher order coherence in swimming behaviour that promotes self-overlap at large scales. Common patterns observed in plankton movements such as spirals clearly fit this picture: in this case the self-overlap quickly increases when the scale is larger than half the pitch of the spiral. Helical swimming is in fact a common mode of locomotion exhibited by plankton including bacteria \cite{Berg1992}, protists \cite{Fenchel1988} and copepods \cite{Titelman2001}.

Plankton trajectories might present more complex shapes that do not fit in the helical pattern but still presenting intriguing regularity. This is the case of \textit{C. furcatus} swimming, which was been considered to be efficient for exploiting micro-patches of food \cite{Uttieri2008, uttieri2013}. However, the observation that \textit{C. furcatus} moves at higher speed compared to other copepod species \cite{Bianco2013} might bring to the misleading conclusion that it is also exposed to high encounter rates with predators. In contrast, the high self-overlap of \textit{C. furcatus} showed at very small scale ($\sim 1$~$mm$) predicts a low exposure to predators. This applies  particularly to the case of a mechanoreceptive predator -- which has a shorter detection radius compared to a visual predator -- and would explain the low predatory rates by chaetognaths on this species \cite{kehayias1996}.

In the study area,  both copepod species have peak abundances in autumn \cite{Mazzocchi2011} and are then likely exposed to similar
environmental conditions as well as similar predation pressure from both mechanoreceptive and visual predators. The different
behaviour in the two species seems to  represent different adaptive modes consequent on evolutionary trade-offs. While \textit{C. furcatus} carries the eggs in a ventral egg-mass, \textit{T. stylifera} is a broadcast spawner and the predation on the former has an higher cost on the individual and species fitness than on the latter. Moreover, \textit{T. stylifera} may be able to perform escape reactions as shown in other congeneric species \cite{van2003escape, waggett2007}, thus showing a post-encounter defence mechanisms that could explain its bolder behaviour.

Any encounter rate estimation is highly dependent on the detection distance that can vary greatly based on the type of prey or predator. However, the method of self-overlap can be used to predict at which scale the detection distance is optimised for different species and different encounter types. It can be ultimately used to design specific experimental setups to further investigate the mechanism of encounter rate in pelagic ecosystems.

From a more theoretical point of view, our results highlight two issues about patterns in zooplankton motion behaviour. First, it is not clear whether a mathematical limit exists to the steepness at which the self-overlap increases, nor as to what the geometry of such optimal trajectories should be. An optimal mathematical model is not easy to implement for 3D trajectories and morphological constrains limit the range of swimming patterns that copepods can perform. A second issue is how to implement movement strategies which are efficient in this sense. Planktonic organisms have a simple neural system and it is implausible that they keep a long-term memory of previously searched regions of space. It is worth exploring how the self-overlap can sharply change in an animal's movement  that responds only to local stimuli, i.e., without storing information about the past trajectory. The aforementioned issues remain open to movement ecologist. The application of the self-overlap metric may be potentially useful in interpreting searching strategy models in a more comprehensive way and could shed light on evolutionary consequences of animal movement.

\section{Methods}

\subsection{Algorithm to compute the effective volume and the self-overlap}

To compute the effective volume, we first fix a maximum radius $r_{max}=10$ $mm$. For each trajectory, we then define an enclosing box, $x_{min}-r_{max}<x<x_{max}+r_{max}$ and similarly for the other coordinates, where $x_{min}$, $x_{max}$ are respectively the minimum and the maximum value of the coordinate $x$ along all the points belonging to the trajectory.

The effective volumes at different values of the radius are then estimated via a Montecarlo scheme \cite{binder2010}: we draw random points whose coordinates are uniformly distributed in the box. For each point, we update the histogram of the number of points falling at a distance less than a given radius $r$ from the trajectories, where the radius $r$ is between $r_{min}=0.05$ $mm$ and the maximum radius defined above. The distance of the random point from the trajectory was measured after linear interpolation between each pair of consecutive trajectory data-points.

At the end of the simulation, the effective volume for each radius is simply calculated as
\begin{equation}
V(r)=\frac{n(r)}{n_{tot}}V_{box}
\end{equation}
where $n(r)$ is the number of points in the histogram at a given $r$, $n_{tot}$ is the total number of points and $V_{box}$ is the volume of the box. In all simulations we fix $n_{tot}=10^8$.

Finally, computing the maximum possible volume $V_{max}(r)$ for a given length as in equation (\ref{V_max}), the scale-dependent self-overlap is given by the definition in equation (\ref{def_selfover}).

When applying the method to the dataset, we checked that the trajectories were long enough so that the results are stationary and not too influenced by the behaviour at the endpoints. Specifically, we randomly selected a few trajectories, divided them into two equally long parts and repeated the analysis on the sub-segments. We verified that the results are compatible with those of the entire trajectory, so that we can safely neglect finite size effects. Only in one case in which there is a clear switch of pattern in the trajectory we observed, as expected, small but appreciable differences in the behaviour of the sub-segments.

\subsection{Trajectories dataset}

\textit{Clausocalanus furcatus} and \textit{Temora stlyfera} specimens were sorted from zooplankton samples collected with a Nansen net (200 $\mu m$ mesh) at the coastal station LTER-MC (40$^{\circ}$48.5$'$N; 14$^{\circ}$15.0$'$E) in the Gulf of Naples (Tyrrhenian Sea, Western Mediterranean). Samples were kept in the laboratory at controlled light and temperature set to \textit{in situ} conditions. Under a dissecting microscope, only healthy and intact adults were selected for recording experiments. Homogeneous groups of the same gender (N=30--37 for \textit{C. furcatus} and N=25 for \textit{T. stylifera}) were moved into a 1-liter cubic glass aquarium and left to acclimatize for 15 minutes in the dark. Food was provided as a natural particle assemblage collected with Niskin bottles at the same time and location of zooplankton sampling. To simulate extreme depleted resources condition, \textit{C. furcatus} was also recorded without food, in GF/F filtered seawater.

The feeding modes differ for the two copepod species. \textit{C. furcatus} was recorded capturing dinoflagellates of $10^{-2}$ $mm$ in size within a distance of $r \approx 0.2$ $mm$ \cite{Uttieri2008}. \textit{T. stylifera} generates feeding currents to capture prey. We could not directly measure the encounter radius of \textit{T. stylifera}, but the volume of influence of the feeding currents for the congeneric species \textit{T. longicornis}, has been estimated to be $12.52$ $mm^3$ \cite{van2003copepod}. While the feeding current is not symmetric around the copepod body, a rough encounter radius $r \approx 1.4$ $mm$ can be estimated assuming a spherical  volume of influence. We consider this radius an approximation for the detection radius of the similar size \textit{T. stylifera}.

Each group of copepods was recorded for several minutes (20--90~$min$) in dark condition using a telecentric stereo vision system as described in Bianco et al. \cite{Bianco2013}. Briefly, the system is based on two Sony XCD-X700 FireWire digital cameras, two telecentric lenses placed orthogonally to each other, and two 780~$\eta m$ infra-red light sources. The system was able to record copepod positions at 15~Hz with 78~$\mu m$ spatial resolution. Custom made software was used to record synchronized videos from both cameras and for 3D tracking. During the tuning of the system it was given particular attention to reduce the positional error of tracked organisms \cite{Bianco2013}. The system was design to record only copepods that are at least 10~$mm$ away from the walls and 20~$mm$ from the bottom or the top of the aquarium to prevent any interference of the walls on the copepod motion. Trajectories of copepods that exit the volume of observation and re-enter successively were recorded as different trajectories. The distance between data points was $0.32 \pm 0.22$ $mm$ for \textit{C. furcatus} and $0.34 \pm 0.16$ $mm$ for \textit{T. stylifera}.

After the recording, the individuals used for the experiments were preserved in a $4\%$ buffered formaldehyde-seawater solution for successive size measurements at a Leica M2 12.5 stereo-microscope. The average total length (from the tip of the prosome to the distal end of the caudal rami) of \textit{C. furcatus} and \textit{T. stylifera} (both genders) was $1.03 \pm 0.04$ $mm$ and $1.45 \pm 0.05$ $mm$, respectively.

Only trajectories longer than 15~$s$ were considered for the present study. A total of 750 trajectories were used for \textit{C. furcatus} and 309 for \textit{T. stylifera} (Table~\ref{tab1}). For \textit{C. furcatus} the trajectories with sinking phases represented 58\% of the total in presence of food, and 89\% without food.

\begin{widetext}
\begin{table}
\caption{General information on trajectories database of \textit{Clausocalanus furcatus} and \textit{Temora stylifera} recorded during 12 experiments. Food is represented by natural particle assemblage sampled at the same site and time of target species. Minimum trajectory duration is 15 s.}
\begin{tabular}{rlcccccc}
\hline
&&&&& \multicolumn{3}{c}{\textit{trajectories duration}} \\
\hline
\it{species} & \it{gender} & \it{food} & \it{nr. of experiments} & \it{nr. of trajectories} & \it{cumulative} & \it{maximum} & \textit{average} \\
\hline
\it{Clausocalanus furcatus} 	& Female & Yes & 2 & 320 & 4h 49 min 22 s & 4 min 38 s & 42 s \\
 						 	&        & No  & 2 & 430 & 6h 19 min 06 s & 5 min 42 s & 54 s \\

\it{Temora stylifera} 		& Female & Yes & 4 & 192 & 2h 18 min 58 s & 4 min 43 s & 52 s \\
 						 	& Male   & Yes & 4 & 117 & 1h 15 min 12 s & 4 min 36 s & 39 s \\
\hline
\end{tabular}
\label{tab1}
\end{table}
\end{widetext}

\begin{acknowledgments}
We are grateful to Thomas Ki{\o}rboe for discussions during the preparation of the manuscript. GB was supported by the Swedish Research Council (VR grant to L.-A.H. 621-2010-5404) and by the Centre for Animal Movement Research (CAnMove) financed by a Linnaeus grant (349-2007-8690) from the Swedish Research Council and Lund University. PM was supported by SUNFISH project and the EU-FP7 grants Euromarine. AWV was supported by NAACOS, a Danish Council for Strategic Research project.
\end{acknowledgments}

\end{document}